\documentclass[aps,prd,twocolumn,showpacs,nofootinbib]{revtex4-1}

\usepackage{graphicx}

\begin{document}

\title{The QCD improved electroweak parameter $\rho$}

\author{Sheng-Quan Wang}

\author{Xing-Gang Wu}
\email[email:]{wuxg@cqu.edu.cn}

\author{Jian-Ming Shen}

\author{Hua-Yong Han}

\author{Yang Ma}

\address{ Department of Physics, Chongqing University, Chongqing 401331, P.R. China}

\date{\today}

\begin{abstract}
In the present paper, we make a detailed analysis for the QCD corrections to the electroweak $\rho$ parameter by applying the principle of maximum conformality (PMC). As a comparison, we show that under the conventional scale setting, we have $\Delta\rho|_{\rm N^3LO} = \left(8.257^{+0.045}_{-0.012}\right) \times10^{-3}$ by varying the scale $\mu_{r}\in[M_{t}/2$, $2M_{t}]$. By defining a ratio, $\Delta R=\Delta\rho/3X_t-1$, which shows the relative importance of the QCD corrections, it is found that its scale error is $\sim \pm9 \%$ at the two-loop level, which changes to $\sim\pm4\%$ at the three-loop level and $\sim \pm 2.5\%$ at the four-loop level, respectively. These facts well explain why the conventional scale uncertainty constitutes an important error for estimating the $\rho$ parameter. On the other hand, by applying the PMC scale setting, the four-loop estimation $\Delta\rho|_{\rm N^3LO}$ shall be almost fixed to $8.228\times10^{-3}$, which indicates that the conventional scale error has been eliminated. We observe the pQCD convergence for the $\rho$ parameter has also been greatly improved due to the elimination of the divergent renormalon terms. As applications of the present QCD improved $\rho$ parameter, we show the shifts of the $W$-boson mass and the effective leptonic weak-mixing angle due to $\Delta\rho$ can be reduced to $\delta M_{W}|_{\rm N^3LO} =0.7$ MeV and $\delta \sin^2{\theta}_{\rm eff}|_{\rm N^3LO}=-0.4\times10^{-5}$.
\end{abstract}


\maketitle

\section{Introduction}

The $\rho$ parameter, being defined as the ratio between the strengths for the charged and neutral currents~\cite{rho1}, plays an important role for the electroweak physics. A precise determination of $\rho$ can further improve the accuracy of the electroweak precision observables (EWPOs), such as it provides strong indirect constraints for the top quark $M_t$~\cite{rhopredi,CDF,D0} and the Higgs mass $M_{H}$~\cite{disin,dimw,higglom}. Thus, it is helpful to derive a more accurate $\rho$ parameter for precision test of standard model (SM) and for finding new physics beyond SM.

The $\rho$ parameter can be schematically written as
\begin{equation}
\rho=1+\Delta\rho \;.
\end{equation}
At the Born level, $\rho|_{\rm Born}=1$, and the shift of $\rho$ caused by loop-corrections can be defined as, $\Delta\rho={\Pi_{Z}(0) / M^{2}_{Z}}-{\Pi_{W}(0) / M^{2}_{W}}$, where $\Pi_{Z}(0)$ and $\Pi_{W}(0)$ are transversal parts of $W$-boson and $Z^0$-boson self-energies at the zero momentum transfer. At present, the one-loop QCD corrections~\cite{rho1}, the two-loop QCD corrections~\cite{rhoqcdtwo1,rhoqcdtwo2,rhoqcdtwo3}, the three-loop QCD corrections~\cite{rhoqcdthr1,rhoqcdthr2,rhoqcdthr3,rhoqcdthr4}, and the four-loop QCD corrections~\cite{rhoqcdfo1,rhoqcdfo2,rhoqcdfo3,rhoqcdfo4} to the $\rho$ parameter have been done in the literature. All those improvements on loop calculations provide us great chances for deriving more accurate QCD estimation for $\rho$.

Under the conventional scale setting, there is renormalization scheme and scale ambiguity for a fixed-order pQCD correction. That is, conventionally, one always takes $\mu_{r}=Q$ ($Q$ being the typical momentum flow of the process) as its central value, and then varies the scale within a certain region, e.g. $\mu_r\in [Q/2, 2Q]$, to ascertain the scale uncertainty. More specifically, we shall show that the conventional scale uncertainty for the $\rho$ parameter is still large even for the four-loop level, thus, it is important to find a reliable way to suppress, or even eliminate, such large scale uncertainty.

The principle of maximum conformality (PMC)~\cite{pmc1,pmc2,pmc3,pmc4,pmc5,pmc6,pmc8,pmc9,pmc11} has been designed for eliminating the renormalization scale ambiguity via a systematic way. By applying the PMC scale setting, all the non-conformal terms in perturbative QCD series are summed into the running coupling, and one obtains a unique, scale-fixed and scheme-independent prediction at any finite order. We shall try to eliminate the renormalization scale ambiguity for $\Delta\rho$ by using the PMC $R_\delta$-scheme~\cite{pmc6,pmc11}. The PMC scales are formed by absorbing the $\{\beta_{i}\}$-terms that govern the behavior of the running coupling via the renormalization group equation into the running coupling. Those $\{\beta_{i}\}$-terms that are related to the quark mass renormalization and etc. should be kept as a separate during the PMC scale setting. To avoid the confusion of using PMC, one can first transform expressions in terms of $\overline{\rm MS}$-quark mass into those of on-shell quark mass~\cite{higrr} and then apply PMC. In the present paper, we shall explain this treatment in detail.

The remaining parts of the paper are organized as follows. In Sec.II, we give our calculation technology for $\Delta\rho$ and show how to deal with it within the framework of PMC. In Sec.III, we present our numerical results for $\Delta\rho$, and also present the application of $\Delta\rho$ for both the shift of the $W$-boson mass $\delta M_{W}|_{\rm N^3LO}$ and the shift of the effective leptonic weak-mixing angle $\delta \sin^2{\theta}_{\rm eff}|_{\rm N^3LO}$ up to four-loop QCD corrections. The final section is reserved for a summary.

\section{Calculation technology for the QCD corrections to the $\rho$ parameter}

For the conventional scale setting, the renormalization scale  $\mu_r$ is fixed to be an initial value $\mu^{\rm init}_r$, which is usually chosen as the typical momentum transfer of the process. While for PMC, the value of $\mu^{\rm init}_{r}$ is arbitrary. Thus, in order to apply PMC properly, we shall first transform the four-loop expression for $\Delta\rho$ derived in Refs.\cite{rhoqcdfo1,rhoqcdfo2, rhoqcdfo3,rhoqcdfo4} into those with full initial scale dependence, in which both the singlet and non-singlet contributions shall be taken into consideration.

Then, we transform the $\Delta\rho$ parameter with the $\overline{\rm MS}$ quark masses into the one with the on-shell quark masses. This transformation is important to separate out the right $\{\beta_{i}\}$-terms that govern the behavior of the running coupling. The relation between the $\overline{\rm MS}$-quark mass and the on-shell quark mass up to three-loop level can be found in Refs.\cite{pomstwo1,pomstwo2,pomstwo3, pomstwo4, pomsthr1, pomsthr2,pomsthr3}. After doing such transformation, all remaining $\{\beta_i\}$-terms are rightly pertained to the running coupling and the PMC scales can be readily determined.

More explicitly, we write done $\Delta\rho$ up to order ${\cal O}(a_s^4)$ in the following:
\begin{widetext}
\begin{eqnarray}
\Delta\rho&=&3X_{t}\bigg[1+c_{1,0}(\mu^{\rm init}_{r})a_{s}(\mu^{\rm init}_{r})+ \bigg(c_{2,0}(\mu^{\rm init}_{r}) +c_{2,1}(\mu^{\rm init}_{r})n_{f}\bigg) a_{s}^{2}(\mu^{\rm init}_{r})+\bigg(c_{3,0}(\mu^{\rm init}_{r})\nonumber\\
&&\quad\quad +c_{3,1}(\mu^{\rm init}_{r})n_{f}
+c_{3,2}(\mu^{\rm init}_{r}) n_{f}^{2}\bigg) a_{s}^{3}(\mu^{\rm init}_{r}) +\mathcal{O}\bigg(a_{s}^{4}\bigg)\bigg],  \label{rhocij}
\end{eqnarray}
\end{widetext}
where $X_{t}=(G_{F}M^{2}_{t})/(8\sqrt{2}\pi^{2})$ stands for the one-loop result~\cite{rho1}, $a_{s}(\mu^{\rm init}_{r})=\alpha_{s}(\mu^{\rm init}_{r})/4\pi$ and $G_{F}$ is the Fermi constant. The coefficients $c_{i,j}(\mu^{\rm init}_{r})$ are put in the Appendix. The $n_{f}$-series in Eq.(\ref{rhocij}) can be unambiguously associated with the $\{\beta_{i}\}$-terms that rightly govern the running behavior of the coupling constant via the $R_\delta$-scheme~\cite{pmc6,pmc11}. In the $R_\delta$-scheme, an arbitrary constant $-\delta$ is subtracted in addition to the standard subtraction $\ln 4 \pi - \gamma_E$ for the $\overline{\rm MS}$-scheme. The $\delta$-subtraction defines an infinite set of new $\overline{\rm MS}$-like renormalization schemes. The $\beta$-function of the coupling constant within any $R_\delta$-scheme is the same as the usual $\overline{\rm MS}$ one. All $R_\delta$-schemes are connected to each other by a scale-displacement relation, e.g. for the schemes with $\delta_1$ and $\delta_2$, their coupling constants are related by
\begin{displaymath}
a_s(\mu_{\delta_1}) = a_s(\mu_{\delta_2}) + \sum_{n=1}^\infty \frac{1}{n!} { \frac{{\rm d}^n a_s(\mu_{r})}{({\rm d} \ln \mu^2_{r})^n} |_{\mu_{r} =\mu_{\delta_2}} (-\delta)^n},
\end{displaymath}
where $\ln\mu^2_{\delta_1}/\mu^2_{\delta_2}=-\delta$. At each perturbative order, the running behavior of the coupling constant is controlled by such displacement relation, which inversely determines the $\{\beta_i\}$-terms that pertain to a specific perturbative order. By collecting up all those $\{\beta_i\}$-terms for the same order, one can obtain the general pattern of non-conformal $\{\beta_i\}$-terms at each perturbative order. More specifically, according to the $R_\delta$-scheme, we can rewrite Eq.(\ref{rhocij}) as
\begin{widetext}
\begin{eqnarray}
\Delta\rho&=&3X_{t}\bigg[1+r_{1,0}(\mu^{\rm init}_{r})a_{s}(\mu^{\rm init}_{r}) + \bigg( r_{2,0}(\mu^{\rm init}_{r})  + \beta_{0}r_{2,1}(\mu^{\rm init}_{r})\bigg) a_{s}^{2}(\mu^{\rm init}_{r})+ \nonumber\\
&& \quad\quad \bigg( r_{3,0}(\mu^{\rm init}_{r}) +\beta_{1}r_{2,1}(\mu^{\rm init}_{r})+2\beta_{0}r_{3,1}(\mu^{\rm init}_{r})+ \beta_{0}^{2}r_{3,2}(\mu^{\rm init}_{r})\bigg) a_{s}^{3}(\mu^{\rm init}_{r})+\mathcal{O}\bigg(a_{s}^{4}\bigg)\bigg], \label{rhorij}
\end{eqnarray}
\end{widetext}
where $\beta_{0} = 11-{2\over 3}n_{f}$, $\beta_{1} = 102-{38\over 3} n_{f}$, and the coefficients $r_{i,j}(\mu^{\rm init}_{r})$ can be derived from the coefficients $c_{i,j}(\mu^{\rm init}_{r})$ defined in Eq.(\ref{rhocij}), which are also put in the Appendix. The $r_{i,0}$ with i=(1,2,3) are conformal coefficients, and the $r_{i,j}$ with $1\leq j\leq i\leq 3$ are non-conformal ones that should be absorbed into the running coupling. After absorbing all those non-conformal terms into the running coupling, we finally obtain the scheme-independent conformal series for $\Delta\rho$, i.e.
\begin{eqnarray}
\Delta\rho&=&3X_{t}\bigg[1+r_{1,0}(\mu^{\rm init}_{r})a_{s}(Q_{1}) + r_{2,0}(\mu^{\rm init}_{r}) a_{s}^{2}(Q_{2}) \nonumber\\
 && + r_{3,0}(\mu^{\rm init}_{r})a_{s}^{3}(Q_{3}) +\mathcal{O}\bigg(a_{s}^{4}\bigg)\bigg]. \label{rhorijPMC}
\end{eqnarray}
Here $Q_{i}$ with $i=(1,2,3)$ are PMC scales. At each perturbative order, there are new types of $\{\beta_i\}$-terms, so we should introduce new PMC scales at each perturbative order so as to absorb all the $\{\beta_i\}$-terms into the running coupling consistently~\cite{pmc8}. The PMC scales $Q_{1}$ and $Q_{2}$ can be written as
\begin{eqnarray}
Q_{1} &=& \mu^{\rm init}_{r}\exp\bigg({1\over 2}{-r_{2,1}+{1\over 2}{\partial\beta\over \partial a_{s}}r_{3,2}\over r_{1,0}-{1\over 2}{\partial\beta\over \partial a_{s}}r_{2,1}}\bigg) , \label{rhorijPMCscale1} \\
Q_{2} &=& \mu^{\rm init}_{r}\exp\bigg(-{1\over 2}{r_{3,1}\over r_{2,0}}\bigg), \label{rhorijPMCscale2}
\end{eqnarray}
where $\beta=-a^{2}_{s}\sum^{\infty}\limits_{i=0} \beta_{i} a^{i}_{s}$. There is no higher-order $\{\beta_{i}\}$-terms to determine $Q_{3}$, we set its value as $\mu^{\rm init}_{r}$. This treatment causes residual scale dependence, which, however, can be highly suppressed \footnote{There is another type of residual scale dependence for the already determined PMC scales, which are smaller and are highly exponentially suppressed~\cite{pmc12}. This is the reason why the large-$\beta_0$ approximation suggested in the literature provides a good approximation for setting the scale in certain processes~\cite{beta0}.}.

\section{Numerical results and discussions}

To do numerical calculation, we take the top-quark pole mass $M_{t}=173.3$ GeV~\cite{toppole}, which is compatible with the $\overline{\rm MS}$ mass $\overline{m}_t(\overline{m}_t)= 163.3$ GeV~\cite{toptt}. The $W$-boson mass $M_{W}=80.385$ GeV and the $Z^0$-boson mass $M_{Z}=91.1876$ GeV~\cite{pdg}.
The Fermi constant $G_{F}=1.16638\times10^{-5}{\rm GeV}^{-2}$. To be consistent, as an estimation of $\Delta\rho$ up to certain QCD loop correction, we will use different $\Lambda_{\rm QCD}$ determined by using world average $\alpha_s(M_{Z})=0.1184$~\cite{pdg}: we use $\Lambda^{(n_f=5)}_{\rm QCD}=0.213$ GeV, and $\Lambda^{(n_f=6)}_{\rm QCD}=0.0904$ GeV for three-loop $\alpha_s$ running; $\Lambda^{(n_f=5)}_{\rm QCD}=0.231$ GeV and $\Lambda^{(n_f=6)}_{\rm QCD}=0.0938$ GeV for two-loop $\alpha_s$ running; $\Lambda^{(n_f=5)}_{\rm QCD}=0.0899$ GeV and $\Lambda^{(n_f=6)}_{\rm QCD}=0.0437$ GeV for the one-loop $\alpha_s$ running.

As a subtle point, as shown by Eqs.(\ref{rhorijPMCscale1},\ref{rhorijPMCscale2}), the PMC scales themselves are in perturbative series, thus the PMC scales shall be improved to a certain degree for the estimation with more and more QCD loop corrections being included. For example, to determine PMC scale $Q_1$ for $\Delta\rho$ up to three-loop level, we have only $\beta_0$ term to determine its value; while for $\Delta\rho$ up to four-loop level, we have both $\beta_0$ and $\beta_1$ terms to determine its value. Within the PMC scale $Q_1$, the $\beta_1$ terms are $\alpha_s$ suppressed in comparison to the leading $\beta_0$ terms and such difference shall further be exponentially suppressed, thus, its value changes slightly.

\subsection{The QCD improved electroweak $\rho$ parameter}

\begin{table}[htb]
\begin{center}
\begin{tabular}{|c|c|c|c|c|c|c|}
\hline
& \multicolumn{3}{c|}{Conventional} & \multicolumn{3}{c|}{PMC} \\
\hline
$\mu^{\rm init}_{r}$ & $M_{t}/2$ & $M_{t}$  & $2M_{t}$  & $M_{t}/2$  & $M_{t}$  & $2M_{t}$ \\
\hline
~$\Delta R_{\rm NLO}$~   & -0.109 & -0.098  & -0.091 &-0.131 & -0.131  & -0.131 \\
\hline
$\Delta R_{\rm N^{2}LO}$ & -0.121 & -0.118  & -0.112 &-0.127 & -0.127  & -0.127 \\
\hline
$\Delta R_{\rm N^{3}LO}$ & -0.124 & -0.123  & -0.118 &-0.126 & -0.126  & -0.126 \\
\hline
\end{tabular}
\caption{Initial scale dependence for the ratio $\Delta R_i$ under the conventional scale setting ($\mu_r\equiv\mu^{\rm init}_r$) and the PMC scale setting, where $i$=NLO, $\rm N^{2}LO$ and $\rm N^{3}LO$ stand for the QCD corrections to the $\Delta\rho$ parameter up to two-loop, three-loop, and four-loop levels, respectively. } \label{tableunscale}
\end{center}
\end{table}

After the PMC scale setting, we obtain a more steady prediction over the scale changes for $\Delta\rho$. To show this point more clearly, we define a parameter:
\begin{equation}
\Delta R_i=\frac{\Delta\rho|_{i}}{3X_{t}} -1 ,
\end{equation}
where $i$=NLO, N$^{2}$LO, N$^{3}$LO stand for the QCD corrections to the $\Delta\rho$ parameter up to two-loop, three-loop, and four-loop levels, respectively. The scale dependence for $\Delta R_i$ under the conventional scale setting and the PMC scale setting are put in Table \ref{tableunscale}, where three typical initial scales $\mu^{\rm init}_{r}=M_{t}/2$, $M_{t}$ and $2M_{t}$ are adopted.

We can see from Table \ref{tableunscale} that under the conventional scale setting, the QCD corrections for $\Delta R_i$ shows a strong dependence on the choice of (initial) scale $\mu^{\rm init}_{r}$. For example, its conventional scale error is $\sim \pm9 \%$ for $\mu^{\rm init}_{r}\in[M_{t}/2$, $2M_{t}]$ at the two-loop level, which changes to $\sim\pm4\%$ at the three-loop level and $\sim \pm 2.5\%$ at the four-loop level. Thus, the scale uncertainty constitutes a systematic error for $\Delta\rho$. In contrast, after applying the PMC scale setting, the value of $\Delta R_i$ is almost unchanged for $\mu^{\rm init}_{r}\in[M_{t}/2$, $2M_{t}]$ even at the two-loop level. In fact, by using the formulas (\ref{rhorijPMCscale1}), (\ref{rhorijPMCscale2}), it is found that the PMC scales themselves are almost fixed, i.e. for any $\mu^{\rm init}_{r}$,
\begin{equation}
Q_1 \simeq 26.2 {\rm GeV} \; {\rm and}\; Q_2\simeq 84.6 {\rm GeV}\;.
\end{equation}
The conformal coefficients $r_{i,0}$, as shown in the Appendix, are also independent of $\mu^{\rm init}_{r}$.

\begin{figure}
\includegraphics[width=0.50\textwidth]{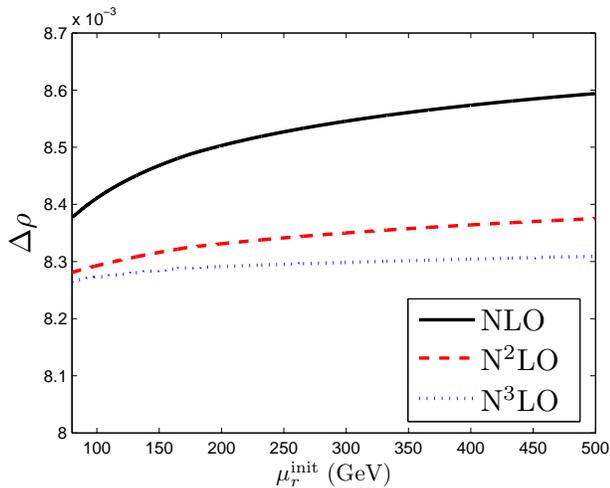}
\caption{The $\Delta\rho$ parameter versus the initial renormalization scale $\mu^{\rm init}_{r}$ under the conventional scale setting.
The solid, dashed and doted lines stand for QCD corrections up to NLO/two-loop, $\rm N^{2}LO$/three-loop and $\rm N^{3}LO$/four-loop, respectively.} \label{Plot:rhoco}
\end{figure}

\begin{figure}
\includegraphics[width=0.50\textwidth]{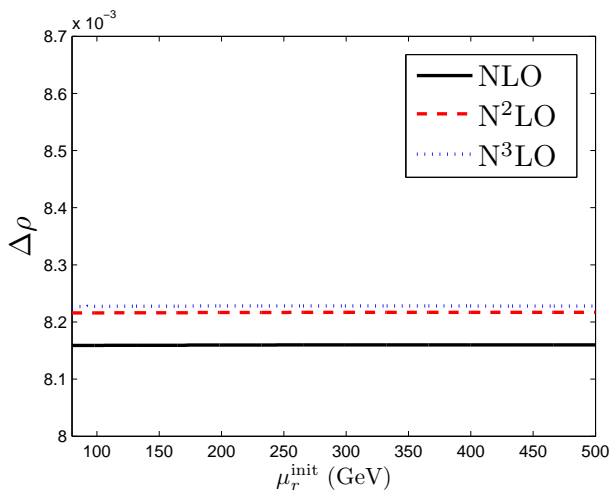}
\caption{The $\Delta\rho$ parameter versus the initial renormalization scale $\mu^{\rm init}_{r}$ under the PMC scale setting.
The solid, dashed and doted lines stand for QCD corrections up to NLO/two-loop, $\rm N^{2}LO$/three-loop and $\rm N^{3}LO$/four-loop, respectively.} \label{Plot:rhopm}
\end{figure}

We present the dependence of $\Delta\rho$ over $\mu^{\rm init}_{r}$ before and after the PMC scale setting in Figs. \ref{Plot:rhoco} and \ref{Plot:rhopm}, where the solid, dashed and doted lines stand for QCD corrections up to NLO/two-loop, $\rm N^{2}LO$/three-loop and $\rm N^{3}LO$/four-loop levels, respectively. Fig.\ref{Plot:rhoco} shows that as one includes higher-and-higher orders, the scale uncertainty will be decreased accordingly to a certain degree: I) By setting $\mu_r=M_t$, we obtain $\Delta\rho\simeq8.49\times10^{-3}$, $8.30\times10^{-3}$, and $8.26\times10^{-3}$ at the two-loop, three-loop and four-loop levels, respectively; II) By setting $\mu_r=M_t/2$, we obtain $\Delta\rho\simeq8.39\times10^{-3}$, $8.27\times10^{-3}$, and $8.24\times10^{-3}$ at the two-loop, three-loop and four-loop levels, respectively. Those results agree with the conventional wisdom that by finishing a higher-order enough calculation, one can finally achieve desirable convergent and scale-invariant estimations. It is often argued that by varying the scale, one can estimate contributions from higher-order terms under the conventional scale setting. However, this procedure can only partly estimate the higher-order contributions, since it only partly exposes the $\{\beta_i\}$-dependent non-conformal terms, not the entire perturbative series~\cite{pmc8}. More explicitly, by varying $\mu^{\rm init}_{r}\in [M_t/2,2M_t]$, Table \ref{tableunscale} shows the central value of $\Delta R_{\rm N^2LO}$ is not within the error of $\Delta R_{\rm NLO}$ and the central value of $\Delta R_{\rm N^3LO}$ is also not within the error of $\Delta R_{\rm N^2LO}$.

On the other hand, after applying the PMC scale setting, we obtain $\Delta\rho\simeq8.17\times10^{-3}$, $8.22\times10^{-3}$, and $8.23\times10^{-3}$ up to two-loop, three-loop and four-loop QCD corrections, respectively. Fig.\ref{Plot:rhopm} shows the $\Delta\rho$ with QCD corrections up to NLO, $\rm N^{2}LO$ and $\rm N^{3}LO$ are almost flat versus the initial scale $\mu^{\rm init}_{r}$. It shows that after the PMC scale setting, the value of $\Delta\rho$ shows a faster steady behavior by including higher-and-higher order corrections, which quickly approaches its steady value with more-and-more loop corrections included. One may even estimate that $\Delta\rho=8.23\times10^{-3}$ could be the final pQCD estimations even by including up to infinite order corrections.  To show how the theoretical prediction changes when more and more loop corrections are included, we define a ratio
\begin{displaymath}
\kappa_{i}=\left|\frac{\Delta R_i - \Delta R_{i-1}}{\Delta R_{i-1}}\right|,
\end{displaymath}
where $i={\rm N^2LO}$, ${\rm N^3LO}$, respectively. This ratio exactly shows how a (`newly') available higher-order correction could be varied from the (`known') lower-order estimation. Under the conventional scale setting, we have
\begin{eqnarray}
& & \kappa_{\rm N^2LO}=11\%,\;\;\kappa_{\rm N^3LO}=2\%\;{\rm for}\;\mu^{\rm init}_{r}=M_t/2 \\
& & \kappa_{\rm N^2LO}=20\%,\;\;\kappa_{\rm N^3LO}=4\%\;{\rm for}\;\mu^{\rm init}_{r}=M_t \\
& & \kappa_{\rm N^2LO}=23\%,\;\;\kappa_{\rm N^3LO}=5\%\;{\rm for}\;\mu^{\rm init}_{r}=2M_t
\end{eqnarray}
While, after the PMC scale setting, we have
\begin{equation}
\kappa_{\rm N^2LO}\simeq 3\%,\;\;\kappa_{\rm N^3LO}\simeq 0.8\%\;{\rm for}\;\mu^{\rm init}_{r}\in[M_t,2M_t].
\end{equation}
Moreover, we note that:
\begin{itemize}
\item The PMC scale at each perturbative order is determined by absorbing particular $\{\beta_i\}$-terms into the running coupling. Other than a guess work for the conventional scale setting, the PMC scales and hence the PMC estimations are highly independent of the choice of initial scale $\mu^{\rm init}_r$. Thus, the conventional renormalization scale ambiguity is solved.

\item A comparison of Table \ref{tableunscale} indicates that after absorbing the non-conformal terms into the running coupling by PMC, the leaving conformal terms shall provide slight positive contributions from the two-loop level, so $\Delta\rho|_i$ shall be increased when more-and-more loop corrections being included. While, under the convention scale setting, the combination of both the conformal and the non-conformal terms shall always provide negative contributions, so $\Delta\rho|_i$ shall decrease with more-and-more loop corrections being included.

\item Another important feature of PMC scale setting is that its final estimation is conformal series and is renormalization scheme independent~\cite{pmc8}, thus the renormalization scale and scheme dependence under the conventional scale setting are eliminated at the same time.

\item For any scale-setting method, we need to finish a full higher-order calculation so as to estimate the magnitude of the conformal terms. The PMC provides a systematic way to estimate the unknown conformal contributions via the extended renormalization group equations~\cite{pmcext}.
\end{itemize}

As a byproduct, from a comparison of Table \ref{tableunscale}, we show that if setting $\mu_r\sim M_t/2$ for the conventional scale setting, one can get the same estimation under the PMC scale setting. Thus, the effective momentum flow for the whole process is $\sim M_t/2$ other than the conventionally suggested $M_t$, or equivalently, it is $\sim M_t/2$ that can rightly eliminate the large logs and get a more convergent/correct pQCD estimation. There are some other examples also show that the conventional choice of scale is really a guess work. Ref.\cite{higrr} indicates that the effective momentum flow for $H\to\gamma\gamma$ decay is $\sim2M_H$ other than $M_H$. Ref.\cite{Q6} argues that after including the first and second order corrections to several deep inelastic sum rules which are due to heavy flavor contributions, the effective scale $\mu_{r}$ for the deep inelastic sum rules should be $\sim 6.5 m_Q$ other than $m_Q$ ($m_Q$ being the heavy quark mass). All those indicate that it is clearly artificial to guess a scale $Q$ (we even do not know whether it is the central scale or not) and to study its uncertainty by simply varying $\mu_{r}\in [Q/2, 2\,Q]$, as the conventional scale setting does.

After the PMC scale setting, there is residual scale dependence for the final terms proportional to $a^3_{s}(Q_{3})$, since we have no $\{\beta_{i}\}$-terms to determine $Q_3$. We can estimate the magnitude of such residual scale dependence following the spirit of PMC scale setting. That is, as suggested in Ref.\cite{jpsi}, we rewrite the coupling constant $a_{s}(Q_{3})$ at the four-loop level as follows,
\begin{eqnarray}
a_{s}(Q_{3})=a_{s}(\mu^{\rm init}_{r})+\beta_{0} \ln\left({(\mu^{\rm init}_{r})^2 \over Q^2_{3}}\right) a^{2}_{s}(\mu^{\rm init}_{r}).  \label{rhorijasq3}
\end{eqnarray}
Since the log term $\ln\left({\mu^{\rm init}_{r}/Q_{3}}\right)^{2}$ can largely compensate the scale changes at the $\mathcal{O}(a_{s}^{3})$ level, and as expected, we obtain a very small residual scale dependence by varying $\mu^{\rm init}_{r}\in[M_t/2,2M_t]$.

\begin{widetext}
\begin{center}
\begin{table}[htb]
\begin{tabular}{|c|c|c|c|c|c|c|c|c|}
\hline
& \multicolumn{4}{c|}{Conventional scale setting} & \multicolumn{4}{c|}{PMC scale setting} \\
\hline
~~~ ~~~ & ~$\rm LO$~ & ~$\rm NLO$~ & ~$\rm N^{2}LO$~&~$\rm N^{3}LO$ ~& ~$\rm LO$~ & ~$\rm NLO$~ & ~$\rm N^{2}LO$~&~$\rm N^{3}LO$ ~ \\
\hline
~$\Delta\rho|_{i} (\times10^{-3})$~ & 9.411 & 8.483  & 8.305 & 8.257 & 9.411 & 8.175 & 8.217 & 8.228 \\
\hline
~$\delta\rho|_{i} (\times10^{-3})$~ & -  &  $-0.928$ & $-0.178$ & $-0.048$ & - & $-1.236$ & $0.042$ & $0.011$ \\
\hline
~$K_{i}$~ & -  &  $9.8\%$ & $2.1\%$ & $0.6\%$ & - & $13\%$ & $0.5\%$ & $0.1\%$ \\
\hline
\end{tabular}
\caption{The parameter $\Delta\rho$, the shift $\delta\rho$, and the $K$ factor before and after the PMC scale setting. $\Delta\rho|_{i}$ with $i$=LO, NLO, N$^{2}$LO and N$^{3}$LO denote the QCD corrections up to one-loop, two-loop, three-loop, and four-loop levels, respectively. The $\delta\rho|_{i}$ and $K_i$ stand for the shift of $\Delta\rho|_{i}$ and $K$ factor for the two-loop, three-loop or four-loop level, respectively. $\mu^{\rm init}_r=M_t$. } \label{tablerho}
\end{table}
\end{center}
\end{widetext}

Finally, we present the QCD correction to $\Delta\rho$ at each perturbative order before and after the PMC scale setting in Table \ref{tablerho}, where $\Delta\rho_{i}$, with $i$=LO, NLO, N$^{2}$LO, or N$^{3}$LO, denote the $\Delta\rho$ with QCD correction up to one-loop, two-loop, three-loop, and four-loop level, respectively. After the PMC scale setting, we obtain a conformal series for $\Delta\rho$, and because of the elimination of renormalons (together with large log-terms), the pQCD convergence can be greatly improved. To show this point more clearly, we define a $K$ factor, whose value at each perturbative order is defined as
\begin{equation}
K_i=\left|\frac{\Delta\rho|_{i}-\Delta\rho|_{i-1}}{\Delta\rho|_{i-1}}\right| =\left|\frac{\delta\rho|_{i}}{\Delta\rho|_{i-1}}\right|,
\end{equation}
where the shift of $\Delta\rho|_{i}$ is defined as $\delta\rho|_{i}= \left(\Delta\rho|_{i}-\Delta\rho|_{i-1}\right)$, and its values are presented in
Table \ref{tablerho}. The $K_i$ stands for the $K$ factor up to two-loop, three-loop or four-loop level, respectively; that is, $i$=NLO, N$^{2}$LO, or N$^{3}$LO, denotes the QCD correction up to one-loop, two-loop, three-loop, and four-loop level, respectively. The results of $K_i$ are also presented in Table \ref{tablerho}. The values for $K$ factors decrease much faster after the PMC scale setting, which agree with the above observation that the pQCD convergence can be greatly improved after PMC scale setting. More over, we obtain $\Delta\rho|_{\rm N^3LO} = \left(8.257^{+0.045}_{-0.012}\right) \times10^{-3}$ for $\mu_r\in[M_t/2,2M_t]$ under the conventional scale setting; at the same time, the $\Delta\rho|_{\rm N^3LO}$ is almost fixed to be $8.228\times10^{-3}$ after the PMC scale setting.

It is noted that several ways to absorb the $\{\beta_i\}$-terms into the running coupling have been suggested, such as the PMC-I approach (based on the PMC-BLM correspondence)~\cite{pmc2}, the PMC $R_\delta$ approach~\cite{pmc6}, and the seBLM approach~\cite{seblm,seblm0}. A detailed comparison of those approaches can be found in Ref.~\cite{higbbgg}. At present, we observe the same conclusions as those of Ref.~\cite{higbbgg}. It is noted that the values of $\Delta\rho$ up to four-loop level agree with each other for those approaches. Because of the choice of different effective $\{\beta_i\}$-series, there are differences at each perturbative order. More explicitly, if setting the initial scale as $\mu^{\rm init}_{r}=M_{t}$, we obtain two effective scales for those approaches as follows
\begin{eqnarray}
R_{\delta}~{\rm approach} : Q_1&=&26.2 {\rm GeV}, Q_2=84.6 {\rm GeV}, \nonumber \\
{\rm PMC-I}\; {\rm approach} : Q_1&=&26.3 {\rm GeV}, Q_2=83.5 {\rm GeV}, \nonumber \\
{\rm seBLM\;approach} : Q_1&=&26.1 {\rm GeV}, Q_2=263.7 {\rm GeV}. \nonumber
\end{eqnarray}
The scales $Q_{1,2}$ are almost the same for both the $R_{\delta}$-approach and the PMC-I approach. Under the seBLM approach, the effective scale $Q_{1}$ is the same as that of the two PMC approaches, but it has a larger $Q_2$.

\subsection{Applications of the QCD improved $\rho$ parameter for $\delta M_W$ and $\delta\sin^{2}\theta^{\rm lept}_{\rm eff}$}

A lot of efforts have been devoted to predict the values of the two important EWPOs as $M_{W}$ and $\sin^{2}\theta^{\rm left}_{\rm eff}$ within SM, either theoretically or experimentally~\cite{unmw,unsin,expersin,ilc}. At present, the total experimental uncertainties for $M_{W}$ and $\sin^{2}\theta^{\rm lept}_{\rm eff}$ are $\delta M_{W} =15$ MeV~\cite{pdg} and $\delta\sin^{2}\theta^{\rm lept}_{\rm eff} =16\cdot10^{-5}$~\cite{expersin}. Recent improvements on Higgs at the LHC~\cite{higgs1,higgs2,higgsmass1,higgsmass2} also allow us to determine the EWPOs with high precision~\cite{elefit1,elefit2,elefit3,elefit4,elefit5}. At the future International Linear Collider (ILC), it is estimated that the accurate experimental uncertainties as $\delta M_{W}=6$ MeV and $\delta\sin^{2}\theta^{\rm lept}_{\rm eff}=13\cdot10^{-5}$ can be achieved~\cite{ilc} \footnote{Through its GigaZ program, $\delta\sin^{2}\theta^{\rm lept}_{\rm eff}$ can be even improved as $1.3\cdot10^{-5}$~\cite{elefit6}.}. On the other hand, theoretically, the dominant shifts of $M_{W}$ and $\sin^{2}\theta^{\rm left}_{\rm eff}$ are due to $\Delta\rho$ though the following formulas~\cite{rhoqcdfo2}
\begin{eqnarray}
\delta M_{W}|_{i} &=& {M_{W}\over 2}{c^{2}_{W}\over c^{2}_{W}-s^{2}_{W}}\delta\rho|_{i} \nonumber\\
&=& {M_{W}\over 2}{c^{2}_{W}\over c^{2}_{W}-s^{2}_{W}} \left(\Delta\rho|_{i}-\Delta\rho|_{i-1}\right) \label{rhoshidm}
\end{eqnarray}
and
\begin{eqnarray}
\delta\sin^{2}\theta^{\rm left}_{\rm eff}|_{i} &=& -{c^{2}_{W}s^{2}_{W}\over c^{2}_{W}-s^{2}_{W}}\delta\rho|_{i} \nonumber\\
&=& -{c^{2}_{W}s^{2}_{W}\over c^{2}_{W}-s^{2}_{W}} \left(\Delta\rho|_{i}-\Delta\rho|_{i-1}\right),  \label{rhoshids}
\end{eqnarray}
where $c_{W}=M_{W}/M_{Z}$ and $s^{2}_{W}=1-c^{2}_{W}$, and sequently, with $i$=NLO, N$^{2}$LO, or N$^{3}$LO, respectively.

\begin{table}[htb]
\begin{center}
\begin{tabular}{|c|c|c|c|c|c|c|}
\hline
& \multicolumn{3}{c|}{Conventional} & \multicolumn{3}{c|}{PMC} \\
\hline
~~~ ~~~ & NLO & $\rm N^{2}LO$  & $\rm N^{3}LO$  & NLO  & $\rm N^{2}LO$  & $\rm N^{3}LO$ \\
\hline
$\delta M_{W}|_{i}$ (MeV) & $-52.3$ & $-10.0$  & $-2.7$ & $-69.7$ & $+2.4$ & $+0.7$ \\
\hline
$\delta\sin^{2}\theta^{\rm left}_{\rm eff}|_{i}(\times10^{-5})$ & $+29.0$ & $+5.6$  & $+1.5$ & $+38.6$ & $-1.3$ & $-0.4$ \\
\hline
\end{tabular}
\caption{The shifts $\delta M_{W}$ and $\delta\sin^{2}\theta^{\rm left}_{\rm eff}$ due to the QCD improved $\rho$ parameter before and after the PMC scale setting, where the symbols NLO, N$^{2}$LO and N$^{3}$LO shifts due to the QCD corrections up to two-loop, three-loop, and four-loop levels, respectively. $\mu^{\rm init}_r=M_t$. } \label{tabledws}
\end{center}
\end{table}

An QCD improved $\rho$ parameter leads to improved estimations on $M_{W}$ and $\sin^{2}\theta^{\rm lept}_{\rm eff}$, which shall help us for a more confidential comparison with the experimental results and for searching new physics beyond SM over those EWPOs. We present the shifts of $M_{W}$ and $\sin^{2}\theta^{\rm left}_{\rm eff}$ caused by the QCD corrections to $\Delta\rho$ before and after the PMC scale setting in Table \ref{tabledws}. At the N$^{3}$LO level, the shifts are $\delta M_{W}|_{\rm N^3LO}$=-2.7 MeV and $\delta\sin^{2}\theta^{\rm left}_{\rm eff}|_{\rm N^3LO}=1.5\times10^{-5}$ under the conventional scale setting, whose precision can be greatly improved by about four times after the PMC scale setting due to a more convergent pQCD series, i.e.
\begin{eqnarray}
\delta M_{W}|_{\rm N^3LO} &=& 0.7 \; {\rm MeV} \label{mw}
\end{eqnarray}
and
\begin{eqnarray}
\delta\sin^{2}\theta^{\rm left}_{\rm  eff}|_{\rm N^3LO} &=& -0.4\times10^{-5} . \label{theta}
\end{eqnarray}

\section{Summary}

We have applied PMC to analyze the electroweak $\rho$ parameter up to four-loop QCD corrections. After the PMC scale setting, we obtain a more accurate estimation on $\Delta\rho$ with a better pQCD convergence, and then a better estimation of the two EWPOs as $\delta M_{W}$ and $\delta\sin^{2}\theta^{\rm lept}_{\rm eff}$ can be achieved. More specifically,
\begin{itemize}
\item We obtain, $\Delta\rho|_{\rm N^3LO} = \left(8.257^{+0.045}_{-0.012}\right)\times10^{-3}$ for $\mu_r\in[M_t/2,2M_t]$ under the conventional scale setting; while, at the same time, the $\Delta\rho|_{\rm N^3LO}$ is almost fixed to be $8.228\times10^{-3}$ after the PMC scale setting. It shows that the conventional scale uncertainty can be eliminated and the pQCD convergence can also be greatly improved by applying the PMC scale setting. Thus, it provides another good example for achieving the optimal renormalization scales of the process via PMC, which has been detailed illustrated in Ref.\cite{pmc5}.

\item In comparison to the results under the conventional scale setting, after applying the PMC scale setting, we obtain a more steady prediction for the shift of $W$-boson mass, $\delta M_{W}$, and the shift of the effective leptonic weak-mixing angle, $\delta\sin^{2}\theta^{\rm lept}_{\rm eff}$. More over, as shown by Eqs.(\ref{mw},\ref{theta}), the QCD improved shifts $\delta M_{W}|_{\rm N^3LO}$ and $\delta\sin^{2}\theta^{\rm left}_{\rm  eff}|_{\rm N^3LO}$ are well below the precision anticipated even for the future ILC experiment. Thus, we shall have great chances to test SM with high precision.

\item To apply the PMC scale setting to higher-order pQCD calculations, it is more convenient to use the expressions of $\Delta\rho$ under the on-shell renormalization scheme, e.g. via using the pole top-quark mass, such that there is no ambiguity in dealing with the $n_f$ series of the process, i.e. only those $n_{f}$-terms that rightly determine the running behavior of the running coupling should be absorbed into the running coupling.
\end{itemize}

\hspace{1cm}

\noindent{\bf Acknowledgments}: This work was supported in part by Natural Science Foundation of China under Grant No.11275280, by the Program for New Century Excellent Talents in University under Grant No.NCET-10-0882, and by the Fundamental Research Funds for the Central Universities under Grant No.CQDXWL-2012-Z002. \\

\section*{Appendix: the coefficients $c_{i,j}(\mu^{\rm init}_{r})$ and $r_{i,j}(\mu^{\rm init}_{r})$}

By using the expressions of Refs.\cite{rhoqcdfo1,rhoqcdfo2,rhoqcdfo3,rhoqcdfo4}, including both the singlet and non-singlet contributions, the coefficients $c_{i,j}(\mu^{\rm init}_{r})$ for $\Delta\rho$ up to four-level are
\begin{eqnarray}
c_{1,0}(\mu^{\rm init}_{r})&=&-{8\over 3}-{8\pi^{2}\over 9},\\
c_{2,0}(\mu^{\rm init}_{r})&=&-404.981+ 125.836 \ln{M^{2}_{t}\over (\mu^{\rm init}_{r})^{2}},\\
c_{2,1}(\mu^{\rm init}_{r})&=&28.5794 - 7.62643 \ln{M^{2}_{t}\over (\mu^{\rm init}_{r})^{2}},\\
c_{3,0}(\mu^{\rm init}_{r})&=&-20372.1 + 10076.4 \ln{M^{2}_{t}\over (\mu^{\rm init}_{r})^{2}} \nonumber\\
&&- 1384.2 \ln^{2}{M^{2}_{t}\over (\mu^{\rm init}_{r})^{2}}, \\
c_{3,1}(\mu^{\rm init}_{r})&=&2843.1 - 1313.62 \ln{M^{2}_{t}\over (\mu^{\rm init}_{r})^{2}} \nonumber\\
&& +167.782 \ln^{2}{M^{2}_{t}\over (\mu^{\rm init}_{r})^{2}},\\
c_{3,2}(\mu^{\rm init}_{r})&=&-73.558 + 38.1059 \ln{M^{2}_{t}\over (\mu^{\rm init}_{r})^{2}} \nonumber\\
&& - 5.0843 \ln^{2}{M^{2}_{t}\over (\mu^{\rm init}_{r})^{2}},
\end{eqnarray}
where $M_t$ stands for the top-quark pole mass.

The coefficients $r_{i,j}(\mu^{\rm init}_{r})$ for $\Delta\rho$ can be divided into two types, one is the conformal type, which includes
\begin{eqnarray}
r_{1,0}(\mu^{\rm init}_{r})&=&-{8\over 3}-{8\pi^{2}\over 9},\\
r_{2,0}(\mu^{\rm init}_{r})&=&66.5793 ,\\
r_{3,0}(\mu^{\rm init}_{r})&=&1925.76,
\end{eqnarray}
and the other is the non-conformal type, which includes
\begin{eqnarray}
r_{2,1}(\mu^{\rm init}_{r})&=&-42.8691 + 11.4396 \ln{M^{2}_{t}\over (\mu^{\rm init}_{r})^{2}},\\
r_{3,1}(\mu^{\rm init}_{r})&=&95.5017 - 66.5793 \ln{M^{2}_{t}\over (\mu^{\rm init}_{r})^{2}},\\
r_{3,2}(\mu^{\rm init}_{r})&=&-165.506 + 85.7382 \ln{M^{2}_{t}\over (\mu^{\rm init}_{r})^{2}} \nonumber\\
&& - 11.4396 \ln^{2}{M^{2}_{t}\over (\mu^{\rm init}_{r})^{2}}.
\end{eqnarray}

\end{document}